# Hybrid AI-Physics Framework for Post-Earthquake Structural Damage Diagnosis with Sparse Sensing

X. Liang[1]

## ABSTRACT

Rapid and reliable post-earthquake damage assessment is critical for public safety, re-occupancy decisions, and effective emergency response. This paper presents a physics-informed, unsupervised learning framework that enables structural damage diagnosis in sparsely instrumented buildings following seismic events. The approach fuses real sensor data with physics-based simulations to create a hybrid spatiotemporal input grid, extending observability to regions without sensors. A Spatiotemporal Composite Autoencoder Network (SCAN) processes this hybrid input, learning the structure's undamaged behavior from pre-event or ambient-condition data alone. SCAN integrates convolutional layers for extracting localized spatial features and LSTM layers for modeling temporal dynamics, enabling it to recognize deviations from normal behavior caused by damage. Post-event sensor data are analyzed through the trained model, and anomalies are flagged based on elevated reconstruction and prediction errors. These error patterns are spatially mapped to localize potential damage, even in uninstrumented areas. By embedding low-fidelity physical estimates directly into the input representation, the framework enhances detection sensitivity while requiring no labeled damage examples. This hybrid AI-physics approach offers a scalable, interpretable, and data-efficient solution for real-time post-earthquake structural diagnostics, providing critical decision support for emergency managers and accelerating safe and targeted recovery efforts.

## Introduction

In November 2018, a magnitude 7.0 earthquake struck near Anchorage, Alaska, causing widespread structural damage. Over 750 buildings were flagged with significant damage, nearly 900 more sustained lesser impacts, and approximately 740 sites remained uninspected weeks later. This backlog strained the city's inspection capacity and delayed recovery. Similar challenges arise after other major disasters – earthquakes, hurricanes, or explosions – where thousands of structures may be affected in minutes, yet post-disaster assessments rely heavily on manual visual inspections like those outlined in FEMA's ATC-20 protocol [1]. These methods are labor-intensive, slow, and often unable to keep pace with the urgency of response, leaving dangerous buildings unchecked and safe ones unnecessarily closed. The consequences include stalled recovery, economic losses, and public safety risks due to missed or delayed diagnoses. To address these limitations, there is an urgent need for automated, scalable, and trustworthy frameworks that can deliver real-time structural assessments under uncertainty. Structural health monitoring (SHM) systems offer promise, but conventional approaches are hampered by sparse sensing coverage and a lack of labeled damage data. In large structures, only a limited number of sensors can be installed, so much of the structure's behavior is unobserved (spatial sparsity). Also, damage labels or examples are exceedingly rare – we often lack data from actual damaged states to train supervised learning models [2-4]. These issues complicate reliable damage detection and localization: damage in an uninstrumented area may go unnoticed, and purely data-driven models may not generalize without labeled damage data. Current model-based methods [5-7] relying on high-fidelity physics models struggle to pinpoint damage with sparse sensors and can be confounded by environmental variations. Pure data-driven approaches like anomaly-detecting autoencoders can flag unusual behavior without labels [8-10], but with few sensors they rarely localize damage precisely. These gaps motivate a new approach that can work unsupervised, leverage physics to cope with sparse sensing, and ultimately improve diagnostic reliability.

[1] Assistant Professor, Zachry Dept. of Civil and Environmental Engineering, Texas A&M University, College Station, TX 77840 (email: xliang@tamu.edu)

To address these challenges, we propose a Hybrid AI-Physics framework for structural damage diagnosis. The core idea is to combine measured sensor data with physics-based simulated data in a unified format that a deep learning model can ingest. Our approach overlays the structure with a grid of points, at which we define three types of inputs: (1) Sensor nodes with real sensor measurements, (2) Virtual sensor nodes with simulated responses from a coarse structural model, and (3) Null nodes outside the structure (no data). We then employ a Spatiotemporal Composite Autoencoder Network (SCAN) to learn the normal behavior of the structure across this grid. Trained only on healthy-condition data, the SCAN can later flag and localize damage by recognizing anomalies in the combined dataset. Crucially, by infusing physics-based estimates at unsensed locations, we "fill in" the structural response field, mitigating blind spots caused by sparse instrumentation. This hybrid AI-physics strategy is novel in SHM: rather than using a physics model only for post-processing or baseline correction, we integrate it directly into the learning input. By doing so, the deep learning model is physics-informed, guiding it with engineering knowledge of how the structure responds, while still allowing it to discover patterns from data.

## Methodology

### Hybrid Input Grid – Fusing Physical and Virtual Sensors

The first component of our framework is the hybrid input data structure (Figure 1). We impose a 2D grid over the structure's geometry (e.g. along the floor plan or elevation), with grid spacing chosen to balance resolution and computational cost. Real sensor nodes on this grid carry measured time-history data (such as accelerations or strains) from the actual sensors installed on the structure. All other grid points within the structural boundaries are designated as virtual sensor nodes, where we generate synthetic responses using a simplified physics model (for instance, a coarse finite-element or modal model of the structure). At each virtual node, the model predicts what the response would be under the current loading and undamaged condition. The grid points lying outside the structure are treated as null nodes – they hold no data and are masked out so as not to influence the model.

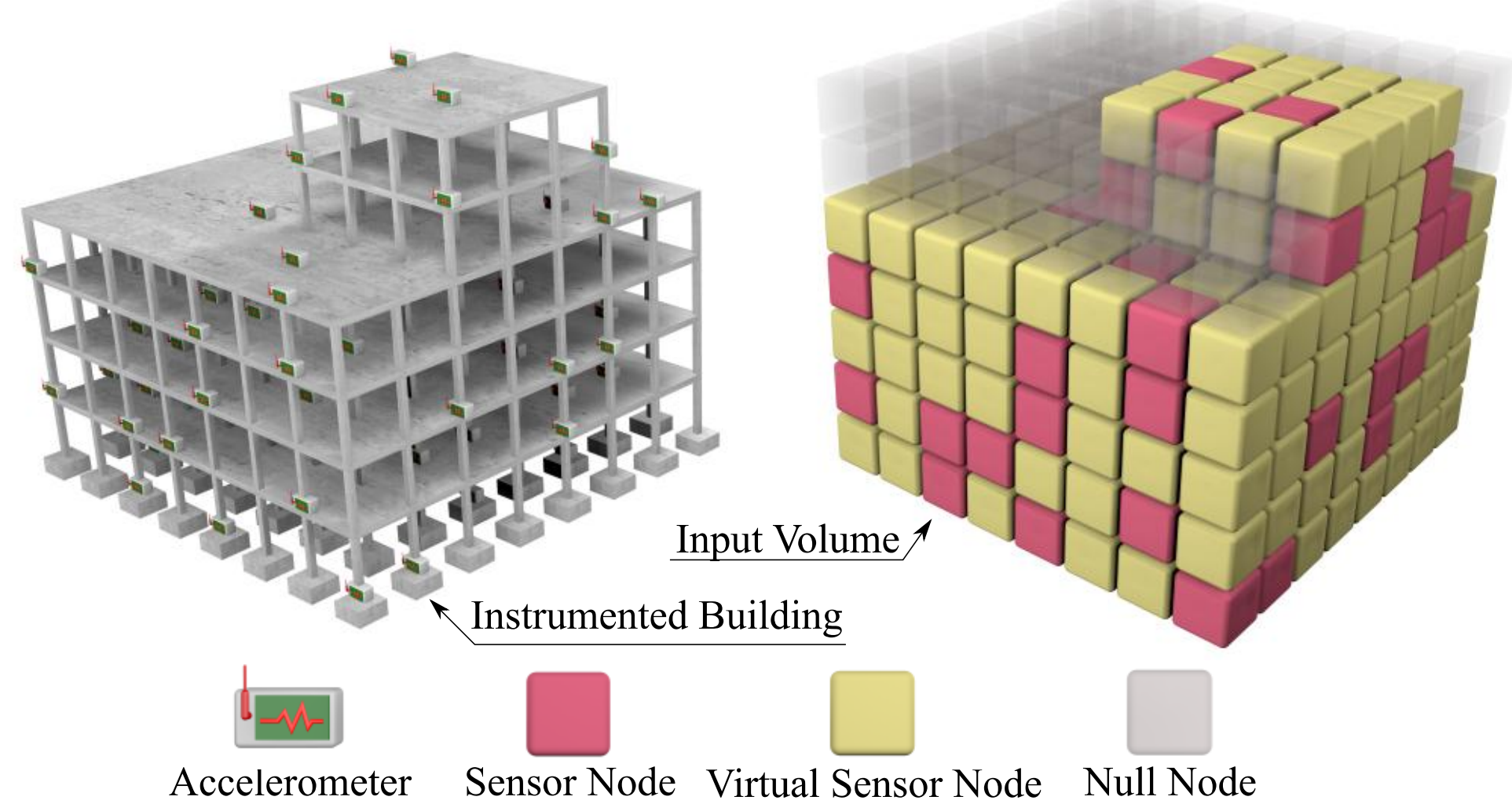

Figure 1. Proposed hybrid sensing grid concept.

All sensor signals (real and virtual) are synchronized in time to form a series of "frames," analogous to images in a video, where each frame is a snapshot of the structural response over the grid at a given time step.

Thus, the hybrid input can be viewed as a spatiotemporal volume of data: two spatial dimensions (grid layout) and one temporal dimension (time series). We apply normalization and masking such that null-node entries do not contribute to training error. In this way, the neural network can focus on learning from the informative portions of the grid (the structure's interior) without distortion from missing data. The fundamental advantage of the hybrid input is improved damage observability: damage at an uninstrumented location will still manifest as an anomaly in the virtual sensor signal at that location, which the network can detect, whereas a conventional approach with no sensor or data there would simply have no information. By blending computational simulation with real data, we inject physics knowledge of the structure's expected behavior into the learning process, effectively regularizing the model and guiding it to interpret structural response patterns. Previous studies have hinted at combining model-based and data-driven methods, but our framework incorporates the physics model at the input level, which is a novel strategy to handle sparse sensing in SHM.

**Spatiotemporal Composite Autoencoder Network (SCAN)**

The SCAN [11,12] is a tailored deep neural network designed to learn the normal structural response from the hybrid input described above. It functions as an unsupervised anomaly detector: trained only on undamaged (baseline) data, it will reconstruct or predict new data, and any significant deviation (error) indicates potential damage. The SCAN architecture (Figure 2) comprises two main parts: a spatial encoder built with convolutional neural networks (CNN) and a temporal encoder using Long Short-Term Memory (LSTM) recurrent units. In the spatial encoder, multiple 2D or 3D convolution layers scan over the grid frames to extract local features. These convolutional filters act on neighboring grid nodes (like pixels), so they are adept at detecting localized changes in the vibration pattern (e.g. a sudden increase in acceleration at adjacent points). Because the conv filters are small and localized, they tend to ignore global shifts that affect all sensors (such as temperature-induced drifts), focusing instead on spatial anomalies (which is exactly what a local damage would produce). Through pooling and successive conv layers, the encoder learns increasingly abstract features, eventually compressing the information into a latent representation (bottleneck). Following the CNN layers, the temporal encoder (LSTM layers) processes the sequence of encoded spatial features across time. The LSTM is critical for modeling the dynamics – it learns temporal patterns such as how a propagating wave or a resonance change unfolds over consecutive time steps. By capturing these spatiotemporal signatures, the SCAN builds a rich internal model of how the undamaged structure behaves. Notably, this composite architecture (CNN + LSTM) means the model can discern transient local anomalies from benign global variations or noise. For instance, an impact or sudden stiffness loss at one location would create a sharp, localized change that the SCAN's filters and memory units should flag, whereas a uniform temperature change would shift all responses slowly and be treated as background.

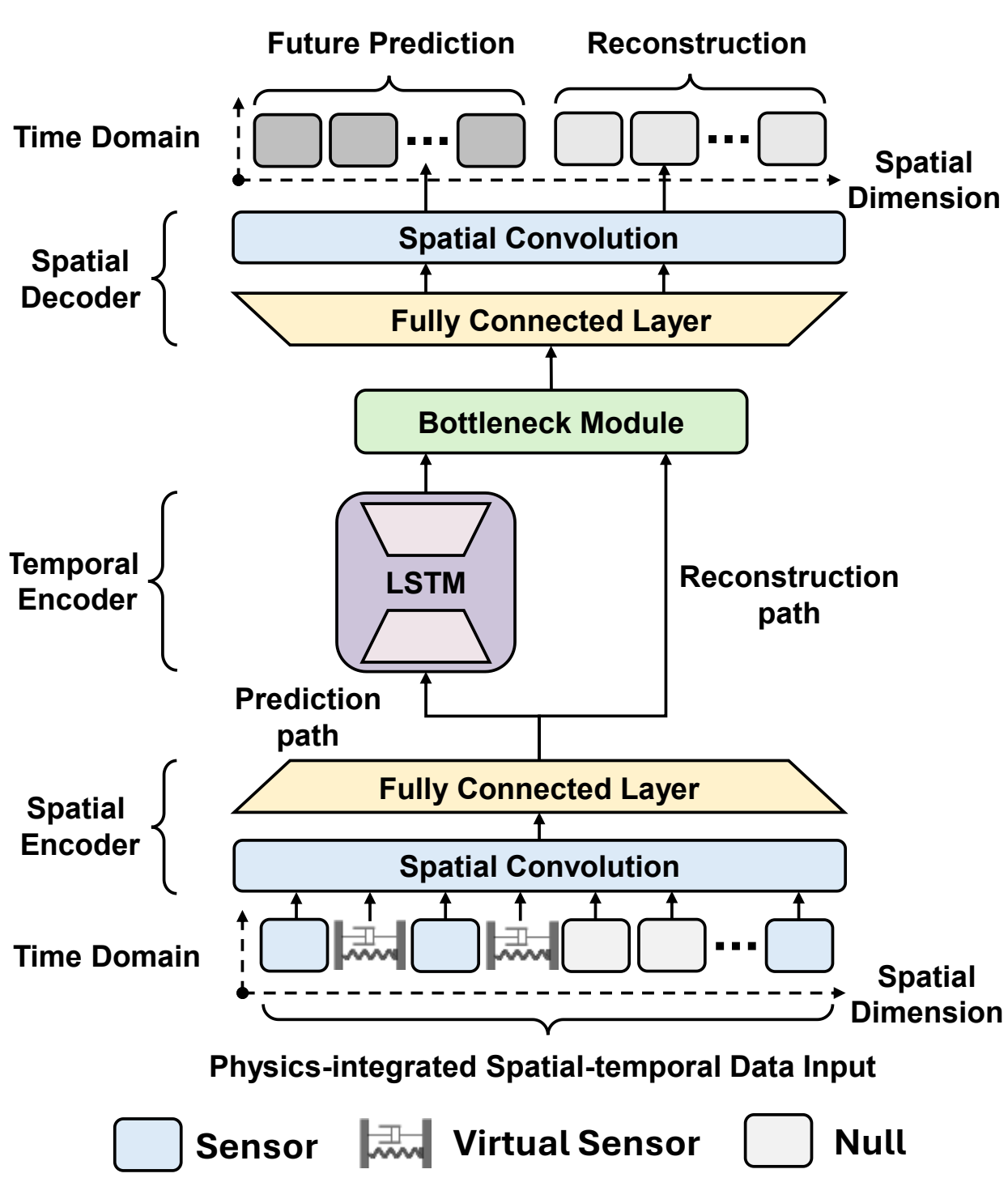

Figure 2. Proposed SCAN architecture.

The SCAN is trained in an unsupervised manner using only normal-condition data. We employ a dual-output training objective: the network is tasked to both reconstruct the input sequence and to predict the next time-step of sensor data. By training for accurate reconstruction, the autoencoder learns to compress the input in a way that preserves important information; by also training to predict future responses, it is encouraged to learn physical dynamics and not just static features. The dual output acts as a consistency check: when the structure is undamaged, the SCAN should both recreate the observed data and foresee its immediate future within a small error. Damage will violate this consistency, leading to elevated reconstruction and prediction errors. During deployment, we feed the SCAN the real-time hybrid grid data. The network outputs a reconstructed "expected" response and a one-step-ahead forecast. Locations and times where the actual measured data differ significantly from the SCAN's output are flagged as anomalies. We aggregate these errors over the grid to localize damage – e.g. a high concentration of error at certain adjacent virtual/real nodes suggests damage in that region. The SCAN's latent features can also be visualized or clustered to provide further insight (though in this paper we focus on the direct anomaly maps). Importantly, our approach does not require any examples of damage for training, only a representative set of healthy operation data. This makes it well-suited to real structures where one cannot obtain labeled damage data in advance. Moreover, by including physics-informed virtual sensors, the SCAN effectively learns to interpolate structural response, propagating information from instrumented points to uninstrumented areas. In summary, the SCAN serves as a high-dimensional novelty detector that is cognizant of structural physics, making it uniquely capable of detecting and roughly localizing damage under real-world sensing constraints.

**Damage Detection and Localization Procedure**

Once the SCAN is trained on the structure's healthy baseline, it can be used for structural diagnostics without retraining. In monitoring mode, we continuously feed the latest sensor readings (augmented with virtual sensor data) into the SCAN and compute the resulting reconstruction error at each grid node and each time step. Sudden increases in reconstruction error (and prediction error) signal that the incoming data do not conform to the learned normal pattern – a strong indication of damage. We implement an anomaly index per grid point, for example the root-mean-square reconstruction error over a moving time window, normalized by the variability under healthy conditions. If this anomaly index exceeds a threshold (set based on training error statistics or a desired false-alarm rate), the point is flagged as damaged. By plotting the anomaly index across all grid locations, we obtain a damage heatmap over the structure. Because the hybrid grid provides estimates everywhere, this heatmap can reveal the damage location: a cluster of high-error nodes suggests the damage's vicinity. The SCAN effectively performs damage localization by anomaly interpolation – even if an actual sensor is not at the damage spot, the nearby virtual sensors (informed by the physics model) will show an error because the real structure's response deviates from the undamaged physics prediction at that location. We expect this hybrid approach to localize damage much more precisely than using sparse real sensors alone. The severity of damage can be gauged qualitatively by the magnitude of the anomaly index or quantitatively by further analysis (e.g. feeding the SCAN's latent features into a regression).

## Conclusions

We presented a physics-informed, unsupervised learning framework for rapid post-earthquake structural damage diagnosis that integrates real sensor measurements with physics-based simulations. Centered on a Spatiotemporal Composite Autoencoder Network (SCAN), the method directly addresses two key challenges in post-disaster structural assessment: sparse instrumentation and the lack of labeled damage data. By constructing a hybrid spatiotemporal input grid that combines measured sensor data with virtual sensor estimates from a coarse physics model, the SCAN learns the structure's undamaged dynamic behavior and flags deviations as potential damage—without requiring any prior examples of damage. This enables wide-area detection and localization even in uninstrumented regions.

The framework's core innovation lies in the seamless fusion of physics and AI. Unlike conventional black-box models, the integration of low-fidelity simulations within the learning process enhances physical interpretability and diagnostic confidence. It also reduces false positives by filtering out benign environmental variability, ensuring that alerts are both targeted and actionable. Because the model is trained only on healthy-state data, it is deployable on real-world structures immediately, with no pre-labeled damage needed. Altogether, this hybrid approach offers a scalable, interpretable, and data-efficient pathway toward real-time, trustworthy post-earthquake diagnostics – supporting faster decisions, safer re-occupancy, and more resilient recovery operations.